\documentclass[twoside,twocolumn,9pt]{article}

\usepackage[utf8]{inputenc}
\usepackage{subfig}
\usepackage{t1enc,latexsym,amsfonts,amssymb,amsmath}
\usepackage{bm}
\usepackage{graphics}
\usepackage{wasysym}
\usepackage{setspace}
\usepackage{url}

\usepackage{extsizes}
\usepackage[super,sort&compress,comma]{natbib}
\usepackage[version=3]{mhchem}
\usepackage[left=1.5cm, right=1.5cm, top=1.785cm, bottom=2.0cm]{geometry}
\usepackage{balance}
\usepackage{times,mathptmx}
\usepackage{sectsty}
\usepackage{graphicx}
\usepackage{lastpage}
\usepackage[format=plain,justification=justified,singlelinecheck=false,font={stretch=1.125,small,sf},labelfont=bf,labelsep=space]{caption}
\usepackage{float}
\usepackage{fancyhdr}
\usepackage{fnpos}
\usepackage[english]{babel}
\usepackage{array}
\usepackage{droidsans}
\usepackage{charter}
\usepackage[T1]{fontenc}
\usepackage[usenames,dvipsnames]{xcolor}
\usepackage{setspace}
\usepackage[compact]{titlesec}
\usepackage{color}

\usepackage{blindtext}

\usepackage{epstopdf}

\usepackage{dcolumn}   

\definecolor{cream}{RGB}{222,217,201}

\begin{document}

\pagestyle{fancy}
\thispagestyle{plain}
\fancypagestyle{plain}{

\fancyhead[C]{\includegraphics[width=18.5cm]{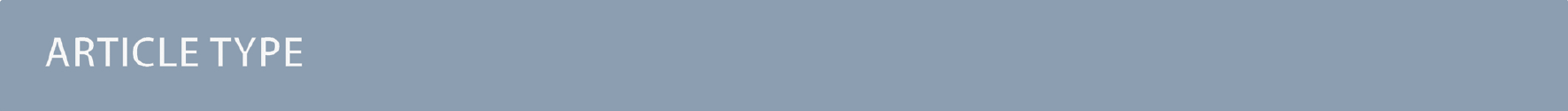}}
\fancyhead[L]{\hspace{0cm}\vspace{1.5cm}\includegraphics[height=30pt]{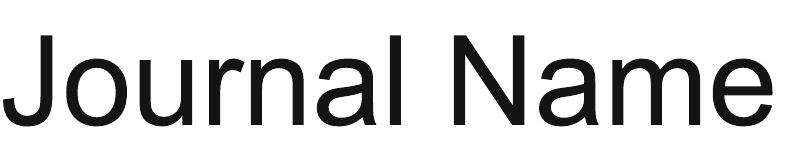}}
\fancyhead[R]{\hspace{0cm}\vspace{1.7cm}\includegraphics[height=55pt]{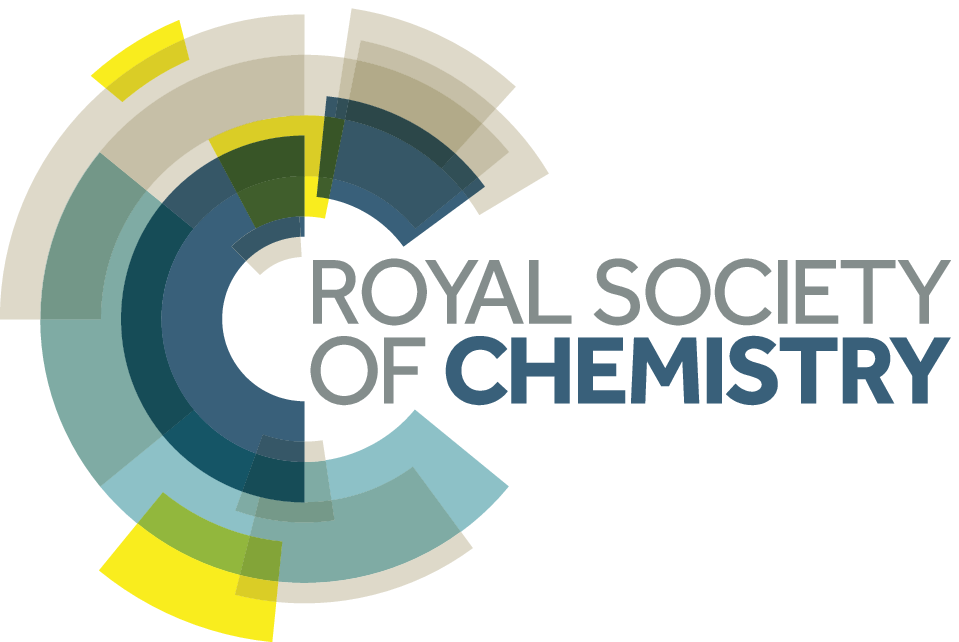}}
\renewcommand{\headrulewidth}{0pt}
}

\makeFNbottom
\makeatletter
\renewcommand\LARGE{\@setfontsize\LARGE{15pt}{17}}
\renewcommand\Large{\@setfontsize\Large{12pt}{14}}
\renewcommand\large{\@setfontsize\large{10pt}{12}}
\renewcommand\footnotesize{\@setfontsize\footnotesize{7pt}{10}}
\makeatother

\renewcommand{\thefootnote}{\fnsymbol{footnote}}
\renewcommand\footnoterule{\vspace*{1pt}%
\color{cream}\hrule width 3.5in height 0.4pt \color{black}\vspace*{5pt}}
\setcounter{secnumdepth}{5}

\makeatletter
\renewcommand\@biblabel[1]{#1}
\renewcommand\@makefntext[1]%
{\noindent\makebox[0pt][r]{\@thefnmark\,}#1}
\makeatother
\renewcommand{\figurename}{\small{Fig.}~}
\sectionfont{\sffamily\Large}
\subsectionfont{\normalsize}
\subsubsectionfont{\bf}
\setstretch{1.125} 
\setlength{\skip\footins}{0.8cm}
\setlength{\footnotesep}{0.25cm}
\setlength{\jot}{10pt}
\titlespacing*{\section}{0pt}{4pt}{4pt}
\titlespacing*{\subsection}{0pt}{15pt}{1pt}

\fancyfoot{}
\fancyfoot[LO,RE]{\vspace{-7.1pt}\includegraphics[height=9pt]{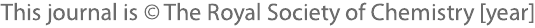}}
\fancyfoot[CO]{\vspace{-7.1pt}\hspace{13.2cm}\includegraphics{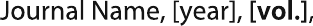}}
\fancyfoot[CE]{\vspace{-7.2pt}\hspace{-14.2cm}\includegraphics{head_foot/RF}}
\fancyfoot[RO]{\footnotesize{\sffamily{1--\pageref{LastPage} ~\textbar  \hspace{2pt}\thepage}}}
\fancyfoot[LE]{\footnotesize{\sffamily{\thepage~\textbar\hspace{3.45cm} 1--\pageref{LastPage}}}}
\fancyhead{}
\renewcommand{\headrulewidth}{0pt}
\renewcommand{\footrulewidth}{0pt}
\setlength{\arrayrulewidth}{1pt}
\setlength{\columnsep}{6.5mm}
\setlength\bibsep{1pt}

\makeatletter
\newlength{\figrulesep}
\setlength{\figrulesep}{0.5\textfloatsep}

\newcommand{\topfigrule}{\vspace*{-1pt}%
\noindent{\color{cream}\rule[-\figrulesep]{\columnwidth}{1.5pt}} }

\newcommand{\botfigrule}{\vspace*{-2pt}%
\noindent{\color{cream}\rule[\figrulesep]{\columnwidth}{1.5pt}} }

\newcommand{\dblfigrule}{\vspace*{-1pt}%
\noindent{\color{cream}\rule[-\figrulesep]{\textwidth}{1.5pt}} }

\makeatother


\newcommand{\argc}[1]{\left[#1\right]}
\newcommand{\arga}[1]{\left\lbrace #1\right\rbrace }
\newcommand{\argp}[1]{\left(#1\right)}
\newcommand{\valabs}[1]{\vert #1\vert}
\newcommand{\moy}[1]{\left\langle  #1 \right\rangle }
\newcommand{\moydes}[1]{\overline{#1}}
\renewcommand{\vec}[1]{{\bf #1}}
\newcommand{\meevid}[1]{\textcolor{VioletRed}{#1}}
\newcommand{\KM}[1]{\textcolor{blue}{#1}}
\newcommand{\comment}[1]{\textcolor{OliveGreen}{#1}}



\twocolumn[
  \begin{@twocolumnfalse}
\vspace{3cm}
\sffamily
\begin{tabular}{m{4.5cm} p{13.5cm} }

\includegraphics{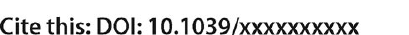} & \noindent\LARGE{\textbf{Non-trivial rheological exponents in sheared yield stress fluids}} \\
\vspace{0.3cm} & \vspace{0.3cm} \\

 & \noindent\large{Elisabeth Agoritsas$^{\ast}$\textit{$^{a,b,c}$} and  Kirsten Martens\textit{$^{b,c}$}} \\
\includegraphics{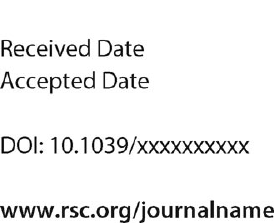} & \noindent\normalsize{
In this work we discuss possible physical origins for non-trivial exponents in the athermal rheology of soft materials at low but finite driving rates. 
A key ingredient in our scenario is the presence of a self-consistent mechanical noise that stems from the spatial superposition of long-range elastic responses to localized plastically deforming regions.
We study analytically a mean-field model, in which this mechanical noise is accounted for by a stress diffusion term coupled to the plastic activity. 
Within this description we show how a dependence of the shear modulus and/or the local relaxation time on the shear rate introduces corrections to the usual mean-field prediction, concerning the Herschel-Bulkley-type rheological response of exponent $1/2$.
This feature of the mean-field picture is then shown to be robust with respect to 
structural disorder and partial relaxation of the local stress.
%
We test this prediction numerically on a mesoscopic lattice model that implements explicitly the long-range elastic response to localized shear transformations,
and we conclude on how our scenario might be tested in rheological experiments.
} \\

\end{tabular}

 \end{@twocolumnfalse} \vspace{0.6cm}

  ]
  


\renewcommand*\rmdefault{bch}\normalfont\upshape
\rmfamily
\section*{}
\vspace{-1cm}


\footnotetext{\textit{$^{*}$~E-mail:} elisabeth.agoritsas@lpt.ens.fr}
\footnotetext{\textit{$^{a}$~Laboratoire de Physique Th{\'e}orique, ENS \& PSL University, UPMC \& Sorbonne Universit{\'e}s, F-75005 Paris, France.}}
\footnotetext{\textit{$^{b}$~Universit\'{e} Grenoble Alpes, LIPHY, F-38000 Grenoble, France.}}
\footnotetext{\textit{$^{c}$~CNRS, LIPHY, F-38000 Grenoble, France.}}




\section{Introduction}
\label{section-intro}

In rheological experiments of athermally driven yield stress materials, such as foams, gels or granular materials, one of the key characterizations is the measurement of the flow curve, i.e.~the relation between the shear stress in the steadily flowing regime and the externally applied shear rate~\cite{bonn_2015_ArXiv-1502.05281}.
In the experimental and numerical literature we find many examples where a Herschel-Bulkley (HB) flow behaviour is observed~\cite{ovarlez_2013_JNonNewtonianFluidMech193_68}.
The steady-state shear stress $\sigma_M$ dependence on the constant shear rate ${\dot{\gamma}}$ is then well-fitted by $\sigma_M = \sigma_y + A_\mathrm{HB} \dot{\gamma}^{\,n}$, where $\sigma_y$ denotes the dynamical yield stress in the zero shear-rate limit and $A_\mathrm{HB}$ the power-law amplitude.
Although the HB exponents have been reported to lie in a range ${n=0.2 \ldots 1}$ (see for example the review on foams~\cite{hoehler_cohen-addad_2005-JPhysCondensMatter41_R1041} and references within), most of the experimental works~\cite{cloitre_2003_PhysRevLett90_068303,becu_2006_PhysRevLett96_138302,katgert_2008_PhysRevLett101_058301,dinkgreve_2015_PhysRevE92_012305, dollet_bocher_2015_EurPhysJE38_123,dinkgreve_2017_RheologicalActa56_189} report on exponents close to or slightly smaller than $1/2$, the exponent that was predicted in a mesoscopic elasto-plastic mean-field model initially introduced by H{\'e}braud and Lequeux~\cite{hebraud_lequeux_1998_PhysRevLett81_2934}.
Interestingly it has been shown in a recent work on carbopol microgels, that this exponent can be related to the exponents describing the fluidization processes in the transient dynamics of shear-rate and stress imposed experiments \cite{divoux_barentin_manneville_2011_SoftMatter7_9335}.


One of the explanations for exponents \emph{smaller} than $1/2$ can be found in inertial effects that are neglected in this mean-field description, and that can nevertheless strongly alter the flow behaviour~\cite{nicolas_barrat_rottler_2016_PRL116_058303, karimi_barrat_2016_PhysRevE93_022904}.
In fact, any process that can introduce a shear banding phenomenon is expected to lower the apparent exponents in the flow curve (see e.g.~Refs.~\cite{mansard_2011_SoftMatter7_5524,
fielding_cates_sollich_2009_SoftMatter5_2378,coussot_2010_EurPhysJE33_83,
martens_bocquet_barrat_2012_SoftMatter8_4197}). 
On the contrary, when considering the overdamped homogeneously flowing regime, it has been shown in numerical simulations on elasto-plastic models that in the close vicinity of zero shear rate, one rather expects \emph{larger} HB exponents than the mean-field predicted one~\cite{lin_2014_PNAS111_14382,liu_2016_PhysRevLett116_065501}.


Regarding the specific case of foams, which are prototypal athermal systems, there have been detailed experimental studies about the relation between the flow curve exponent and the microscopic properties of the foam membranes~\cite{princen_kiss_1989_JColloidInterfaceSci128_176}.
For instance, it has been shown that the HB exponent can be related to the bubble surface mobility~\cite{denkov_2009_SoftMatter5_3389} and can vary between 
$n=0.2\ldots0.5$.
A visco-elastic theory of Schwartz and Princen~\cite{schwartz_princen_1987_JColloidInterfaceSci118_201} suggests in particular an exponent of $n=2/3$ for the rheology of foams, however it does not take into account the elasto-plastic picture that includes the long-range elastic reponse to T1 events~\cite{cohen-addad_hoehler_2014_CurrentOpinionColloidInterfaceScience19_536}.
A complete theory for foams should obviously take into account all of the above effects including both the local properties of the dissipation process in the foam films as well as the elasto-plastic description.

\begin{figure}[!t]
\includegraphics[width=0.49 \textwidth]{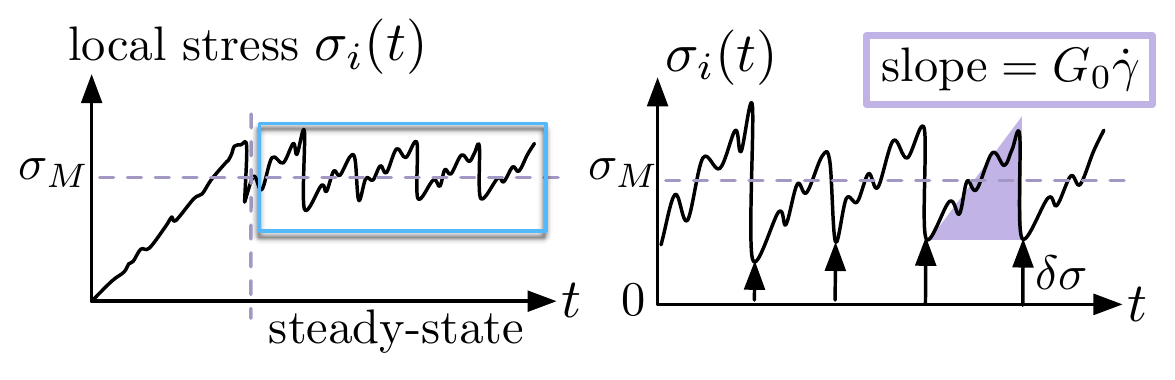}
\caption{
Schematic evolution of the local shear stress, coarse-grained on a scale given by the typical size of a T1 event.
\textit{Left:}~Transient and steady-state regimes.
\textit{Right:}~Zoom on the steady-state fluctuating stress; the typical local slope ${\partial_t \sigma_i(t)}$ inbetween two plastic events defines the steady-state shear modulus $G_0$, and the local slope is actually distributed around its average ${G_0 \dot{\gamma}}$; after each plastic event, the local stress has partially relaxed to a value $\delta \sigma$ which is distributed, as explicitly included in \eqref{eq-def-HL-PDF}.}
\label{fig:stress-strain}
\end{figure}

In this work we show, using an analytically amenable mean-field model based on the elasto-plastic scenario and relevant for intermediate shear-rate regimes~\cite{Puosi_Olivier_Martens_2015_Softmatter11_7639}, that not only the specific choice of the fitting range of $\dot{\gamma}$, but also shear-rate dependent elastic and dissipative properties can effectively change the value of the apparent exponents.
This is particularly relevant since shear and loss moduli have indeed been reported to be frequency-dependent, for example in the literature on foams in oscillatory driving experiments on emulsion, foams and gels~\cite{mason_weitz_1995_PhysRevLett74_1250, liu_1996_PhysRevLett76_3017, gopal_2003_PhysRevLett91_188303, piau_2007_JNon-NewtonianFluidMech144_1, dollet_raufaste_2014_CRPhysique15_731,rouyer_2008_EPJE27_309}.
In situations with steadily applied shear, one can expect this frequency dependence to translate to a rate dependence of the local dissipation process and of the steady-state shear modulus $G_0$ in the elasto-plastic flow regime (see sketch in Fig.~\ref{fig:stress-strain}). 
Accordingly, the aim of this work is twofold:
\textit{(i)}~to assess further the robustness of the ${1/2}$ HB behaviour prediction~\cite{agoritsas_2015_EPJE38_71}, that appears to be omnipresent in athermal rheology experiments;
\textit{(ii)}~to show how small deviations from the $1/2$ exponent are easily introduced even within a very minimal mean-field picture, when its effective parameters turn out to be shear-rate dependent.


The outline of this paper is the following.
After defining a generic class of mean-field models for athermal sheared materials,
we discuss how a shear-rate dependence of the shear modulus ${G_0(\dot{\gamma})}$ and/or of a typical time scale for the plastic events ${\tau (\dot{\gamma})}$ can introduce corrections to the scaling of the flow curve.
Then we investigate the robustness of the ${1/2}$ HB scaling in the regime of driving rates where a diffusive noise in the local stresses remains a valid assumption.
We compare these predictions to numerical simulations of a spatially-resolved lattice model,
and we conclude on the validity of our elasto-plastic scenario for rheological experiments.

\section{Mean-field models for athermal dynamics}
\label{section-defALYSmodels}

We start by defining a generic class of models for sheared amorphous materials, based on an athermal local yield-stress (ALYS) criterion for the stress dynamics inspired by the Hébraud-Lequeux (HL) model~\cite{hebraud_lequeux_1998_PhysRevLett81_2934}.
To simplify the problem we consider an incompressible medium and we assume that the orientation of the local plastic events corresponds to the macroscopically imposed shear direction,
allowing for a scalar description of the shear stress~\cite{nicolas_2014_SoftMatter10_4648}.
We define the class of ALYS models with the following ingredients, in a coarse-grained picture for the stress dynamics in a shear-rate-controlled protocol:
\textit{(i)}~Regions deform elastically according to a rate $G_0 \dot{\gamma}$,
with $G_0$ the local shear modulus.
\textit{(ii)}~When the local stress $\sigma$ exceeds the local stress threshold ${\sigma_c>0}$ (that can be distributed throughout the system), a region `yields', i.e.~it deforms plastically, with a rate ${\nu(\sigma,\sigma_c)}$.
\textit{(iii)}~Local shear stress fluctuations are described by a stochastic process that depends on the surrounding plastic activity.
As we have argued in Ref.~\cite{agoritsas_2015_EPJE38_71} in a toy-model picture, such ALYS models should be suited for describing \emph{athermal} systems, and they encompass in fact several existing models~\cite{hebraud_lequeux_1998_PhysRevLett81_2934,
bocquet_PhysRevLett103_036001,
picard_2005_PhysRevE71_010501,
nicolas_martens_barrat_2014_EurPhysLett107_44003,
agoritsas_2015_EPJE38_71,
jagla_2015_PhysRevE92_042135,
lin_wyart_2016_PhysRevX6_011005,
bouchaud_2016_SoftMatter12_1230}.
Such a stress-threshold-based dynamics has to be distinguished from thermally activated process for which some energy barrier has to be overcome~\cite{sollich_1997_PhysRevLett78_2020,sollich_1998_PhysRevE58_738, nicolas_martens_barrat_2014_EurPhysLett107_44003}; here local barriers disappear when the stress threshold is reached.


In particular, introduced as a minimal mean-field description of the athermal rheology of soft glasses,
the original HL model~\cite{hebraud_lequeux_1998_PhysRevLett81_2934}
encodes the mechanical noise through a \emph{diffusion process of the local shear stress}, whose coefficient is proportional to the overall plastic activity triggered by the external shear rate $\dot{\gamma}$.
This model predicts that, in the steady state at low shear rate,
we can have different types of flow behaviours depending on the value of the coupling parameter $\alpha$ between the diffusion coefficient and the plastic activity. 
For $\alpha$ smaller than a limiting value ${\alpha_c=\sigma_c^2/2}$, the system displays a HB flow behaviour with a $1/2$ scaling of the macroscopic stress: ${\sigma_M \approx \sigma_Y + A_{\text{HB}} \, \dot{\gamma}^{1/2}}$.
Both the macroscopic yield stress ${\sigma_Y(\alpha,\sigma_c)}$ and the prefactor ${A_{\text{HB}}(\alpha,\sigma_c)}$ depend on the coupling and the typical local yield stress, however the main feature to retain for now is the exponent $1/2$.
Note that, in thermally activated processes as in Ref~\cite{sollich_1997_PhysRevLett78_2020, sollich_1998_PhysRevE58_738}, the HB exponent would on the contrary depend on the effective `temperature' $x$ controlling the Arrhenius escape rates, being specifically given by $(1-x)$, providing thus a completely different prediction and physical origin of the HB exponent.
And the well-studied `shear-transformation-zone' theory on the other hand predicts a Bingham fluid in the low shear-rate limit instead of Herschel-Bulkley type flow \cite{langer_2015_PhysRevE92_012318}.

\begin{figure}[!t]
\includegraphics[width=0.45 \textwidth]{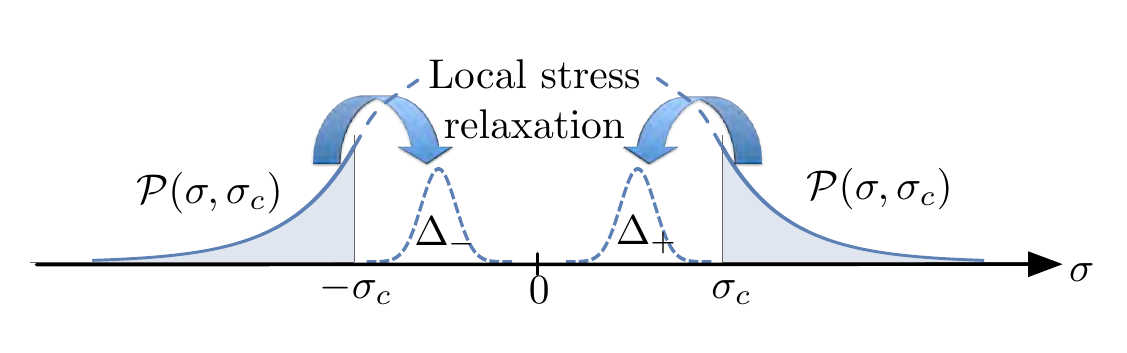}
\includegraphics[width=0.49 \textwidth]{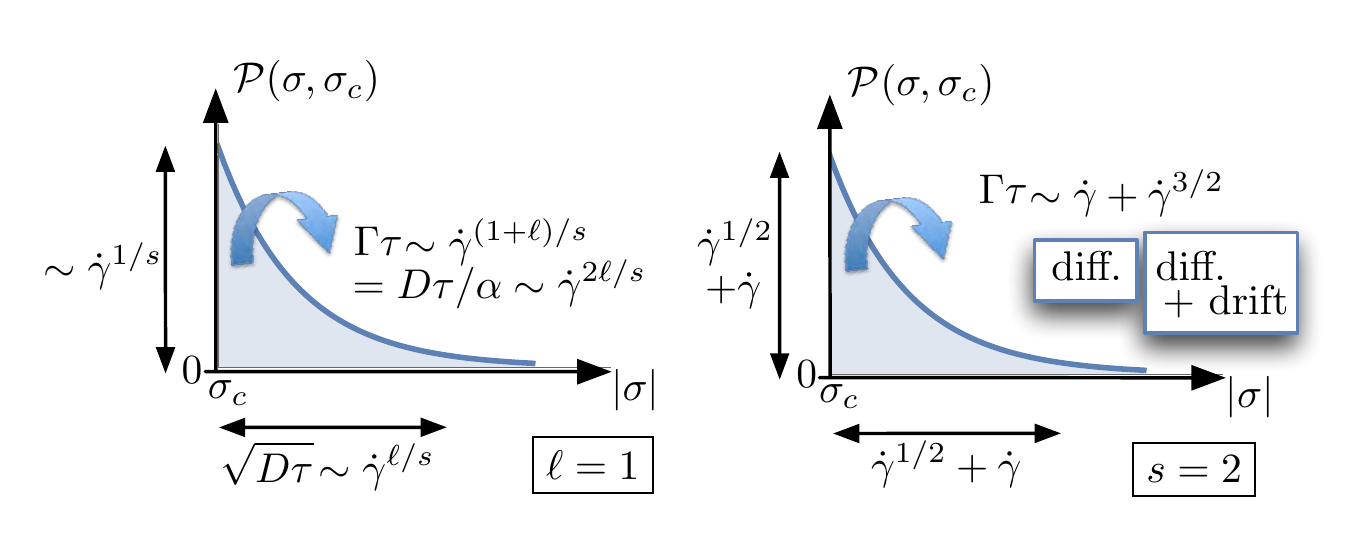}
\caption{
\textit{Top:}~Sketch of the stress redistribution in \eqref{eq-def-HL-PDF}:
yielding sites (${\valabs{\sigma}>\sigma_c > \sigma_c^\mathrm{min}}$) must be disconnected from  just relaxed ones (${\valabs{\sigma}<\sigma_c^\mathrm{min}}$).
\textit{Bottom:}~Scalings of the limiting layer of overstressed sites for ${\alpha < \alpha_c}$, emphasizing the competing contributions of the diffusion and the shear-rate drift
(here $G_0 \tau$ are fixed so ${\Sigma_{\dot{\gamma}} \sim \dot{\gamma}}$).
}
\label{fig:HL-PDF-sketch}
\end{figure}

Inspired by the HL model, we can thus define more broadly the \emph{diffusive} ALYS models, combining the following mean-field ingredients:
\textit{(a)}~We assume a fixed plastic rate for overstressed sites, explicitly ${\nu(\sigma,\sigma_c)= \frac{1}{\tau} \theta(\valabs{\sigma} - \sigma_c)}$, where ${\theta}$ is the Heaviside function.
\textit{(b)}~After a local rearrangement we draw randomly a new local yield stress according to the \textit{a priori} distribution ${\rho (\sigma_c)}$, and relax immediately the local stress to a value $\delta \sigma$ (see sketch in Fig.~\ref{fig:stress-strain}) somewhere below both the previous and this new local threshold.
\textit{(c)}~We assume diffusive stress fluctuations with a coefficient $D(t)$ proportional to the plastic activity ${\Gamma (t)=\langle \nu (\sigma,\sigma_c) \rangle = \int d\sigma_c \int d \sigma \, \nu (\sigma,\sigma_c)}$, i.e.~with the closure relation ${D(t) = \alpha \Gamma(t)}$~\cite{bocquet_PhysRevLett103_036001,agoritsas_2015_EPJE38_71}.
Combining these ingredients, we can then write down the following evolution equation of the probability distribution function (PDF) of local stress $\sigma$ and local yield stress $\sigma_c$, at a time~$t$:
\begin{equation}
\begin{split}
 \partial_t \mathcal{P}(\sigma, \sigma_c , t)
 = & - G_0 \dot{\gamma} \, \partial_\sigma \mathcal{P}
 	+ D(t) \, \partial_\sigma^2 \mathcal{P}
 	- \nu(\sigma,\sigma_c) \,  \mathcal{P} \\
 	& + \Gamma_+(t) \, \Delta_+(\sigma,\sigma_c)  \rho (\sigma_c) \\
 	& + \Gamma_-(t) \, \Delta_-(\sigma,\sigma_c) \, \rho (\sigma_c)
\end{split}
\label{eq-def-HL-PDF}
\end{equation}
where the plastic activities $\arga{\Gamma_{+},\Gamma_{-}}$ are given by the proportion of overstressed sites 
$\Gamma_{\pm} (t) 
= \pm\frac{1}{\tau} \int_0^{\infty} d\sigma_c \int_{\pm \sigma_c}^{\pm \infty} d \sigma \, \mathcal{P}(\sigma,\sigma_c,t)$,
and the total plastic activity is self-consistently ${\Gamma(t)=\Gamma_+(t)+\Gamma_-(t)}$.
The residual stress after these rearrangements is characterized by the distributions ${\Delta_{\pm}(\sigma,\sigma_c) \geq 0}$ with a finite support ${\valabs{\sigma}<\sigma_c^{\mathrm{min}}}$,
without any overlap with the yield stress distribution ${\rho (\sigma_c)}$ defined on a support ${\valabs{\sigma}\geq\sigma_c^\mathrm{min}}$
as illustrated in Fig.~\ref{fig:HL-PDF-sketch}.
The arbitrary residual stress distributions are expected to be symmetric with respect to the yielding direction with ${\Delta_-(\sigma,\sigma_c)=\Delta_+(-\sigma,\sigma_c)}$.
At last, the different distributions are normalized as
${\int_{-\sigma_c^\mathrm{min}}^{\sigma_c^\mathrm{min}} d \sigma \, \Delta_{\pm} (\sigma,\sigma_c) = 1}$
and ${\int_{\sigma_c^\mathrm{min}}^{\infty} d \sigma_c \, \rho (\sigma_c) = 1}$.
This is illustrated in Fig.~\ref{fig:stress-strain}, for the local stress itself.


Assuming ${\Delta_{\pm}(\sigma,\sigma_c)=\delta(\sigma)}$, a full local stress relaxation, we recover either the original HL model~\cite{hebraud_lequeux_1998_PhysRevLett81_2934} with a unique local yield stress (${\rho(\sigma_c)=\delta(\sigma_c-\sigma_c^*)}$), or its disordered generalisation~\cite{agoritsas_2015_EPJE38_71} with an extended ${\rho (\sigma_c)}$.
In the disordered case,
the distinction between the local yield stress before and after a local rearrangement is important; 
indeed, the local yielding from ${(\sigma,\sigma_c)}$ to ${(\sigma',\sigma_c')}$ can be described in full generality by a distribution ${\widetilde{\Delta}(\sigma, \sigma';\sigma_c,\sigma_c')}$.
Nevertheless, we assumed implicitly in \eqref{eq-def-HL-PDF} that
${\widetilde{\Delta}(\sigma, \sigma';\sigma_c,\sigma_c')=\Delta_{\text{sgn}(\sigma')}(\sigma',\sigma_c') \,\rho (\sigma_c')}$,
i.e.~that the final local stress and yield stress have no memory of their values prior to the plastic rearrangement.


The validity of the behaviour predicted by such diffusive ALYS models depends \textit{a fortiori} on the shear-rate regime for which such a diffusive mean-field approximation is a relevant description for the fluctuations of the local shear stress in a sheared athermal system.
In fact, the evolution equation \eqref{eq-def-HL-PDF} can be reformulated with a stochastic equation ${\partial_t \sigma (t) = G_0 \dot{\gamma} + \xi_{\text{pl}}(t)}$
with ${\xi_{\text{pl}}(t)}$ a mechanical noise accounting for the propagation of stress from the surrounding boxes experiencing a local relaxation of stress,
coupled to a resetting dynamics when ${\valabs{\sigma}>\sigma_c}$ with a fixed rate $\tau$ and a plastic rearrangement characterized by ${\Delta_{\pm}}$ and ${\rho(\sigma_c)}$.
One interpretation of the diffusion assumption in \eqref{eq-def-HL-PDF} is that ${\xi_{\text{pl}}(t)}$ can be approximated by a Gaussian white noise of variance ${2D(t)}$,
thus neglecting the time correlation of the mechanical noise.
However, it has been argued on the contrary that, too close to the critical point of zero shear rate, a mean-field description should assume that $\xi_{\text{pl}}(t)$ has the typical power-law distribution of Lévy flights~\cite{lin_wyart_2016_PhysRevX6_011005}.
Nevertheless, the Gaussian and uncorrelated noise approximations should be recovered when increasing the shear rate, because of the increasing density of local rearrangements and the decorrelation of the stress noise signal, according to the scenario for critical exponents characterizing the yielding transition and the avalanche dynamics recently presented in Ref.~\cite{liu_2016_PhysRevLett116_065501}.
Indeed, in this study, the rheological exponent has been shown to cross over from $0.65$ to $0.51$ as departing from the critical point of zero shear rate, approaching thus consistently the $1/2$ HB prediction of the mean-field model.


We emphasize that the study presented thereafter is relevant \emph{within the stress diffusion assumption},
which is at the core of several previous models~\cite{hebraud_lequeux_1998_PhysRevLett81_2934,
bocquet_PhysRevLett103_036001,
mansard_2011_SoftMatter7_5524,
agoritsas_2015_EPJE38_71,
jagla_2015_PhysRevE92_042135,
bouchaud_2016_SoftMatter12_1230}.
Our work is thus complementary to these previous studies, and it is self-contained within this diffusive scenario.
Moreover, the quantitative comparison between molecular dynamics simulations and the HL model has been addressed in Ref.~\cite{Puosi_Olivier_Martens_2015_Softmatter11_7639}, discussing both the assumptions and the predictions of the HL model with respect to flow curves of sheared bidisperse Lennard-Jones mixtures, which display a HB behaviour compatible with an exponent $1/2$.
At last, the specific Kinetic-Elasto-Plastic (KEP) model introduced in Ref.~\cite{bocquet_PhysRevLett103_036001} has provided an analytical derivation of the HL model itself and its closure relation ${D=\alpha \Gamma}$,
as well as of the phenomenological `non-local fluidity equation', which has in turn been successfully applied to study experimental systems such as in Refs.~\cite{goyon_2010_SoftMatter6_2668,
kamrin_koval_2012_PhysRevLett108_178301,
bouzid_2015_EPJE38_1}.
So, although the diffusive assumption is restricted to an intermediate regime of low but finite shear rates, 
these previous studies and their connections with experiments support furthermore the relevance of investigating analytically the class of diffusive ALYS models,
as we do thereafter.
More refined features of the rheology such as the exponent $\theta$ characterizing the density of shear transformations, discussed for instance in Ref.~\cite{lin_wyart_2016_PhysRevX6_011005}, are known not to be correctly captured by diffusive ALYS models such as the HL model, but they are beyond the scope of the present work centered on the averaged mean stress of the flow curve.


\section{Shear-dependent effective parameters}
\label{section-sheardependent-effective-param}

Having motivated the choice of our analytical solvable model, we now highlight and discuss the first main result of our work, i.e.~the dependence of the flow curve exponent with respect to shear-rate dependent elastic and dissipative properties in the stationary state.
To do so, we first recall the main results in the usual case, where such a possible dependence is neglected.
In the steady state and for $\alpha$ below a limiting coupling strength $\alpha_c$, the diffusive ALYS models predict the following scalings, for the stress diffusion coefficient $D$ and the macroscopic stress~$\sigma_M$:
\begin{equation}
\begin{split}
 D\tau 
 & \stackrel{(\dot{\gamma} \to 0)}{\approx}
 C_1 \, G_0 \dot{\gamma} \tau \argc{1 + C_2 \, (G_0 \dot{\gamma} \tau)^{1/2}} 
 \\
 \sigma_M (\dot{\gamma})
 & \stackrel{(\dot{\gamma} \to 0)}{\approx}
 \sigma_Y + A \, \argp{G_0 \dot{\gamma} \tau}^{1/2}
 = \sigma_Y + A_{\text{HB}} \dot{\gamma}^{1/2}
\end{split}
\label{eq-scaling-HL-diffusion-stress}
\end{equation}
with $\sigma_Y$ and ${A}$ depending only on $\alpha$ and $\sigma_c$ in the original HL model~\cite{olivier_renardy_2011_SIAMJApplMath71_1144,Puosi_Olivier_Martens_2015_Softmatter11_7639},
and also on the \textit{a priori} distribution of local threshold ${\rho (\sigma_c)}$ in its disordered variant~\cite{agoritsas_2015_EPJE38_71}.
We will establish later on, analytically, the robustness of the $1/2$ HB exponent for a generic diffusive ALYS model at low shear rates.
Nevertheless, a key point is that, due to the structure of the evolution equation \eqref{eq-def-HL-PDF}, all the dependence on $\dot{\gamma}$ is actually on the effective parameter ${\Sigma_{\dot{\gamma}}=G_0 \dot{\gamma} \tau}$, naturally introduced when the evolution equation is written in an adimensional form.
Physically, the quantity ${\Sigma_{\dot{\gamma}}}$ is the typical stress elastically accumulated on the time scale $\tau$, according to the elastic modulus $G_0$.


Many rheological experiments in athermal yield stress fluids exhibit exponents that are close to $1/2$ HB scaling, but not exactly equal~\cite{hoehler_cohen-addad_2005-JPhysCondensMatter41_R1041,cloitre_2003_PhysRevLett90_068303}.
%
%
One natural way to explain deviations from the $1/2$ scaling within the diffusive scenario is to take into account a possible effective shear-rate dependence of the shear modulus $G_0$ and/or the typical time scale of the plastic events $\tau$.
The low shear-rate perturbative expansions \eqref{eq-scaling-HL-diffusion-stress}, whose variant derivations are detailed in  Refs.~\cite{olivier_renardy_2011_SIAMJApplMath71_1144,olivier_2010_ZAngewMathPhys61_445,Puosi_Olivier_Martens_2015_Softmatter11_7639,agoritsas_2015_EPJE38_71},
remain in fact valid as long as ${\Sigma_{\dot{\gamma}}=G_0 \dot{\gamma} \tau \to 0}$ when ${\dot{\gamma} \to 0}$.


For instance, if we assume the following power-law dependence of the elastic modulus and the typical duration of events in the considered shear-rate regime (note that such power laws would of course be expected to have at most a lower cutoff)
\begin{equation}
 G_0(\dot{\gamma})
 \approx
 g_0 \, \dot{\gamma}^{\psi_1}
 \, , \quad
 \tau(\dot{\gamma})
 \approx
 \tau_0 \, \dot{\gamma}^{-\psi_2}
\end{equation}
we can predict that, if ${(\psi_2 - \psi_1) \in \left[ 0,1 \right)}$, we have
\begin{equation}
\begin{split}
 \sigma_M (\dot{\gamma})
 & \approx
 \sigma_Y + A \, g_0 \tau_0 \, \dot{\gamma}^{(1+\psi_1-\psi_2)/2}\;,
\end{split}
\label{eq-scaling-HL-diffusion-stress-effective}
\end{equation}
hence a HB behaviour with the non-trivial exponent ${n=(1+\psi_1-\psi_2)/2}$ instead of ${n=1/2}$.
Note that the power-law dependence is chosen here to illustrate the most simple scenario of altering the flow curve exponent.
But on the other hand it seems also natural to think about a power-law dependence of the shear modulus as a result of a visco-elastic timescale competing with the inverse shear-rate, and a power-law dependence of the typical duration of plastic events due to a competition between a relaxation time and the external driving.


The exponent for the shear-rate dependence of the elastic modulus in the elasto-plastic regime has been chosen positive with respect to results on the dependence of the shear modulus in oscillatory experiments on emulsions, foams and gels \cite{mason_weitz_1995_PhysRevLett74_1250, liu_2016_PhysRevLett116_065501, gopal_2003_PhysRevLett91_188303, piau_2007_JNon-NewtonianFluidMech144_1,dinkgreve_2017_RheologicalActa56_189}.
We expect that a frequency dependence in the steady state of oscillatory experiments translates into a dependence of the shear modulus in the elastic parts of the elasto-plastic flow regime (see sketch of Fig.~\ref{fig:stress-strain}).
In this analogy the role of the corresponding amplitude that should be considered for the oscillatory case should be played by the typical strain accumulated locally in the elastic parts of the flow regime.
It has been shown for emulsions and foams that the dependence of the shear modulus in the low frequency regime is existent but rather weak \cite{mason_weitz_1995_PhysRevLett74_1250, liu_2016_PhysRevLett116_065501, gopal_2003_PhysRevLett91_188303, piau_2007_JNon-NewtonianFluidMech144_1}.
On the other hand we expect the typical dissipation time of events to be more sensitive to the value of the applied shear rate and the exponent should have the opposite sign.
The faster the flow, the stronger the rearrangements will be interrupted through the external drive \cite{petit_2015_JFluidMech763_286}.

Alternatively, for a given rheological experiment or simulation, if we have access to the complete shear-rate dependence of the local quantities ${G_0(\dot{\gamma})}$ and/or ${\tau(\dot{\gamma})}$, we can plot the macroscopic stress as a function of ${\Sigma_{\dot{\gamma}} = G_0(\dot{\gamma}) \, \dot{\gamma} \, \tau(\dot{\gamma})}$.
If the elasto-plastic scenario is the dominant one and the diffusive assumption is valid on the considered shear-rate regime, we can expect to recover the $1/2$ HB behaviour for $\sigma_M(\Sigma_{\dot{\gamma}})$.

\section{Robustness of the ${1/2}$ HB exponent}
\label{section-robustnessHB12}

In order to complete the argument we have just presented, we now analytically assess the robustness of the $1/2$ HB scaling of the flow curve at low effective shear rates, for $\alpha$ below a coupling strength $\alpha_c$ in diffusive ALYS models, defined with respect to distributed yield stresses and local partial stress relaxation.
These two ingredients are present in real experiments, so it is necessary to determine their implications for the HB prediction for it to be experimentally relevant.


In the steady state at constant shear rate, relevant for predicting the flow curve, defining as before ${\Sigma_{\dot{\gamma}} = G_0 \dot{\gamma} \tau}$ (which we recall has the dimension of a \emph{stress}),
the equation \eqref{eq-def-HL-PDF} becomes:
\begin{equation}
\begin{split}
 0
 = & - \Sigma_{\dot{\gamma}} \, \partial_\sigma \mathcal{P}
 	+ D\tau  \partial_\sigma^2 \mathcal{P}
 	- \theta(\valabs{\sigma}-\sigma_c) \,  \mathcal{P} \\
 	& + \tau \, \Gamma_+  \Delta_+(\sigma,\sigma_c)  \rho (\sigma_c)
 	 + \tau \, \Gamma_-  \Delta_-(\sigma,\sigma_c) \, \rho (\sigma_c)
\end{split}
\label{eq-def-HL-PDF-steadystate}
\end{equation}
On the one hand, integrating these different contributions over the understressed sites ${\valabs{\sigma}<\sigma_c}$,
we obtain the following constraint:
\begin{equation}
 \Gamma \tau \, \rho(\sigma_c)
 = \underbrace{\argc{\Sigma_{\dot{\gamma}} \, \mathcal{P} (\sigma,\sigma_c)}_{\sigma=-\sigma_c}^{\sigma=\sigma_c}}_{\text{shear-rate drift}}
 	- \underbrace{D \tau \, \argc{\partial_\sigma \mathcal{P} (\sigma,\sigma_c)}_{\sigma=-\sigma_c}^{\sigma=\sigma_c}}_{\text{stress diffusion}}
\label{eq-def-HL-PDF-steadystate-continuity}
\end{equation}
Physically, this relation simply makes explicit the balance reached in the steady state, between the sites drifted towards their local yield stress by the external shear rate and the diffusion due to the surrounding mechanical noise that can slow down their pace, so that the plastic activity can be characterized solely by the PDF and its derivative at ${\sigma = \pm \sigma_c}$.
On the other hand, the PDF decays exponentially at ${\valabs{\sigma}>\sigma_c}$ and at low shear rate the depletion of these overstressed regions is controlled at lowest order by the diffusion coefficient:
${\mathcal{P}(\sigma,\sigma_c) \propto e^{-(\valabs{\sigma}-\sigma_c)/\sqrt{D\tau}}}$.
Although the specific relaxation ${\Delta_{\pm}}$ will modify the explicit solution of the PDF,
it will not intervene in the $1/2$ HB scaling at low shear rate, which will be fixed by the combination of 
\textit{(i)}~the balance relation \eqref{eq-def-HL-PDF-steadystate-continuity},
\textit{(ii)}~the closure relation ${D=\alpha \Gamma (D)}$,
and \textit{(iii)}~the diffusive exponential decay of the PDF at ${\valabs{\sigma} \gtrsim \sigma_c}$ at lowest order in the shear rate.


At that point, we can rely on the methodology framework for studying the ${\Sigma_{\dot{\gamma}}\to 0}$ limit, detailed in Ref.~\cite{olivier_renardy_2011_SIAMJApplMath71_1144,olivier_2010_ZAngewMathPhys61_445} and summarized in Ref.~\cite{Puosi_Olivier_Martens_2015_Softmatter11_7639} for the original HL model.
In the absence of shear, either the coupling $\alpha$ is sufficiently large so that there can be a self-sustained plastic activity for a finite portion of overstressed sites, or the coupling is too weak and ${\lim_{\Sigma_{\dot{\gamma}} \to 0} \mathcal{P}(\sigma,\sigma_c)=0}$ at ${\valabs{\sigma}>\sigma_c}$.
The limiting value $\alpha_c$ is defined by the transition between these two cases.
The mere possibility of a finite self-sustained activity in the limit of a vanishing shear rate
would be of course non-physical in the absence of any other external driving and/or energy input to the system.
Using an energetic argument on the original HL model, it can be shown that in the regimes ${\alpha \geq \alpha_c}$ 
all the energy released by the plastic events is redistributed by stress diffusion, without any loss.
%
Thus, only our regime of interest at ${\alpha < \alpha_c}$ does describe a realistic stress dynamics with energy dissipation.


From now on, the strict mathematical limit ${\Sigma_{\dot{\gamma}}\to 0}$ should be physically interpreted as a range of sufficiently low, but finite shear rates at which
the assumption of diffusive stress fluctuations is valid. In this regime, there is a limiting layer of overstressed sites at ${\valabs{\sigma} \gtrsim \sigma_c}$, as sketched in Fig.~\ref{fig:HL-PDF-sketch} (\textit{bottom}), which vanishes in the limit of small shear rate in the case ${\alpha < \alpha_c}$, by definition of $\alpha_c$.
This suggests for that case the following ansatz for a Taylor expansion of the PDF~\cite{olivier_renardy_2011_SIAMJApplMath71_1144,olivier_2010_ZAngewMathPhys61_445},
first for the overstressed sites (${\valabs{\sigma}>\sigma_c}$):
\begin{equation}
\begin{split}
 \mathcal{P}(\pm \valabs{\sigma},\sigma_c)
 &= \sum_{k=1}^{\infty} \Sigma_{\dot{\gamma}}^{k/s} \, R_{\pm}^{(k)}\argp{\frac{\valabs{\sigma}-\sigma_c}{\Sigma_{\dot{\gamma}}^{\ell/s}}}
 \\
 &= \Sigma_{\dot{\gamma}}^{1/s} \, R_{\pm}^{(1)}\argp{\frac{\valabs{\sigma}-\sigma_c}{\Sigma_{\dot{\gamma}}^{\ell/s}}}
	+ \mathcal{O} \argp{\Sigma_{\dot{\gamma}}^{2/s}}
\end{split}
\end{equation}
and secondly for the understressed sites (${\valabs{\sigma} \leq \sigma_c}$) as
${\mathcal{P}(\sigma,\sigma_c) = \sum_{k=0}^{\infty} \Sigma_{\dot{\gamma}}^{k/s} \, Q^{(k)}(\sigma,\sigma_c)}$
with ${Q^{(0)}(\pm \sigma_c,\sigma_c)=0}$ and ${s,\ell \in \mathbb{N}}$.
The only needed assumptions for this ansatz is
\textit{(i)}~that there exists a regular perturbative expansion of the PDF in ${\Sigma_{\dot{\gamma}}^{k/s}}$ with ${k \in \mathbb{N}}$, and
\textit{(ii)}~that for overstressed sites the PDF extends on a vanishing layer scaling in ${\Sigma_{\dot{\gamma}}^{\ell/s}}$. 
Matching order by order the PDF and its derivative at the boundaries ${\sigma=\pm \sigma_c}$,
the so-called `matched asymptotic expansion' presented in Refs.~\cite{olivier_renardy_2011_SIAMJApplMath71_1144,olivier_2010_ZAngewMathPhys61_445},
the contributions to the balance equation \eqref{eq-def-HL-PDF-steadystate-continuity} can be identified and interpreted physically order by order.
In our case of interest ${\alpha <\alpha_c}$, illustrated in Fig.~\ref{fig:HL-PDF-sketch} (\textit{bottom}), at lowest order the plastic activity (${\sim \Sigma_{\dot{\gamma}}^{(1+\ell)/s}}$) stems exclusively from the diffusive exponential decay due to the mechanical noise (${\sim \Sigma_{\dot{\gamma}}^{2/s}}$), and thus imposes ${\ell=1}$.
The next order contribution to the plastic activity (${\sim \Sigma_{\dot{\gamma}}^{(2+\ell)/s}}$) results from the balance between the diffusion (${\sim \Sigma_{\dot{\gamma}}^{3/s}}$) and the stress drift due to the external shear rate (${\sim \Sigma_{\dot{\gamma}}^{1+2/s}}$), imposing ${s=2}$.
In fact, beyond giving a recipe for the systematic construction of the complete perturbative expansion, this low shear-rate argument is rather generic,
since it hides the details of the specific local relaxation ${\Delta_{\pm}}$ into the PDF coefficients
${\lbrace Q^{(k)}, R_{\pm}^{(k)}\rbrace}$ at ${\sigma=\pm \sigma_c}$.
The macroscopic stress as a function of the external shear rate is obtained at last by computing order by order the mean value of the steady-state PDF, using moreover the linearity in $\sigma_c$:
\begin{equation}
\begin{split}
 & \sigma_M (\Sigma_{\dot{\gamma}})
 = \int_0^{\infty} d \sigma_c \argp{\int_{\valabs{\sigma} \geq \sigma_c} + \int_{\valabs{\sigma} < \sigma_c}} d \sigma \, \sigma \, \mathcal{P}(\sigma,\sigma_c)
 \\
 &= \underbrace{\sigma_Y \argc{Q^{(0)},\rho(\sigma_c)}}_{\text{macrosc. yield stress}} + \underbrace{A \argc{Q^{(1)},\rho(\sigma_c)}}_{\text{HB prefactor}} \, \Sigma_{\dot{\gamma}}^{1/2}
	+ \mathcal{O}(\Sigma_{\dot{\gamma}})
\end{split}
\label{eq-sigmaM-perturbative-expansion}
\end{equation}
thus recovering as expected the prediction of a HB behaviour with an exponent $1/2$ at low shear rate.
We have given the explicit expressions of $\sigma_Y$ and $A$ for the disordered HL model with full relaxation in Ref.~\cite{agoritsas_2015_EPJE38_71}, and a procedure for computing them directly in the low shear-rate limit of the original HL model is summarized in Ref.~\cite{Puosi_Olivier_Martens_2015_Softmatter11_7639} and could \textit{a priori} be applied to the extended model defined in eqs.~\eqref{eq-def-HL-PDF}-\eqref{eq-def-HL-PDF-steadystate}.
Physically, Eq.~\eqref{eq-sigmaM-perturbative-expansion} means that the scalings of the mean stress are controlled by the scalings of the vanishing layer of \emph{overstressed} sites in the PDF --~fixed in the steady state by the closure relation~-- which in turn control the \emph{understressed} sites PDF.


We emphasize that this prediction does not depend on the details of the local relaxation ${\Delta_{\pm}}$,
and that the scaling of the limiting layer of overstressed sites is self-consistently fixed by the closure relation ${D=\alpha \Gamma (D)}$.
The existence of the three flow-curves regimes in $\alpha$ is generic, the $1/2$ HB scaling at ${\alpha < \alpha_c}$ is robust, the Newtonian regime at ${\alpha > \alpha_c}$ as well, but the limiting scaling at ${\alpha = \alpha_c}$ might be modified by the local relaxation.
Indeed, on the one hand, we have just discussed the robustness of the low-shear-rate perturbative expansion at ${\alpha < \alpha_c}$ rooted in the self-consistent scaling of the vanishing layer of overstressed sites.
On the other hand, in the regime ${\alpha > \alpha_c}$
the model allows for a self-sustained plasticity and hence a finite layer of overstressed sites even in absence of shear rate, with in particular 
${Q^{(0)}(\pm \sigma_c,\sigma_c)>0}$, leading to
${ D\tau \sim \Sigma_{\dot{\gamma}}^{0}}$ and the Newtonian behaviour ${\sigma_M \sim \Sigma_{\dot{\gamma}}}$.
As for the limiting regime ${\alpha = \alpha_c}$, assuming a full relaxation of stress as in the original HL model~\cite{hebraud_lequeux_1998_PhysRevLett81_2934}, it corresponds to having both ${Q^{(0)}(\pm \sigma_c,\sigma_c)=Q^{(1)}(\pm \sigma_c,\sigma_c)=0}$,
leading eventually to ${\lbrace \ell=2, s=5 \rbrace}$, ${D\tau \sim \Sigma_{\dot{\gamma}}^{4/5}}$, and  ${\sigma_M \sim \Sigma_{\dot{\gamma}}^{1/5}}$.


Beyond this low-shear-rate regime, the original HL model predicts a crossover towards a Newtonian regime  (i.e.~linear in $\dot{\gamma}$)~\cite{hebraud_lequeux_1998_PhysRevLett81_2934,agoritsas_2015_EPJE38_71}, a phenomenon that is usually not observed in HB-type fluids~\cite{dinkgreve_2015_PhysRevE92_012305}.
This discrepancy is probably an artefact of the simplified yielding rules in the mean-field description that fail to describe the strong driving regime. 
%
With these specific rules one expects a departure at higher shear rates from the $1/2$ HB behaviour, due to higher order terms in shear rate in the development of the steady state stress and a crossover to the linear regime, which is bounded from below by the limiting behaviour at $\alpha=\alpha_c$ (${\sigma_M \propto \dot{\gamma}^{1/5}}$ for the HL model with a full relaxation of the stress).
In particular, the closer $\alpha$ is to $\alpha_c$, the larger the shear-rate range over which this crossover extends.
This means that, when we tune the range of shear rates ${[\dot{\gamma}_{\text{min}},\dot{\gamma}_{\text{max}}]}$ for the HB fit, we might change the measured exponent already within the original HL model (representative of diffusive ALYS models):
 \textit{(i)}~if $\dot{\gamma}_{\text{max}}$ is too large, the fit will include both the crossover and the Newtonian regime, so the effective exponent $n$ will be overestimated and close to $1$;
 \textit{(ii)}~if $\dot{\gamma}_{\text{max}}$ is small, we fit as expected the exponent $1/2$;
 \textit{(iii)}~the closer to $\alpha_c$, the more restricted this ideal fitting range because of the increasing influence of the power-law behaviour of exponent $1/5$, so we might even obtain an exponent $m$ \emph{smaller} than $1/2$.
We emphasize that these different scaling regimes arise \emph{within the diffusive assumption}, because of the closure relation of the stress diffusion coefficient and the plastic activity which induces a nonlinear shear-rate dependence of ${D(\dot{\gamma})}$, thus limiting the validity range of the $1/2$ regime.


With respect to sheared athermal materials one might thus expect to observe a $1/2$ exponent in the HB flow curve on a limited range of intermediate shear rates. Still, experiments
on emulsion for example show that this regime might extend to the whole range of accessible shear rates \cite{princen_kiss_1989_JColloidInterfaceSci128_176}, which seems to indicate that the Gaussian approximation of the mechanical noise (and hence the diffusive ALYS model) remains valid and the true critical regime at zero shear rate~\cite{liu_2016_PhysRevLett116_065501} cannot be accessed in these materials.
So the robustness of the HB scalings that we have discussed here, that is actually restricted to an intermediate range of shear rates, should match the ranges accessible experimentally (or in simulations) for ${\Sigma_{\dot{\gamma}}=G_0 \dot{\gamma} \tau}$, in order to allow for such a comparison.
This implies in particular that it is crucial to examine the power-law exponent error-bars with respect to the shear-rate fitting range, when analysing a given set of data.

\section{Flow curves analysis for a spatial elastoplastic model}
\label{section-meso-simulations}

In the following, we probe the former ideas on a lattice model that takes into account a time-resolved local yielding process and the spatially-resolved elastic response to local plastic rearrangements~\cite{picard_2005_PhysRevE71_010501, nicolas_martens_barrat_2014_EurPhysLett107_44003, liu_2016_PhysRevLett116_065501}.
To address the same type of athermal dynamics as the ALYS models we restrict our focus to athermal local yielding criteria.
Furthermore, to keep the model as simple as possible, we implement a scalar description and neglect convection effects.
These simplifications have been shown not to influence significantly the macroscopic flow response~\cite{nicolas_2014_SoftMatter10_4648}.


We map the mesoscopic dynamics of the yield stress material onto a lattice model, where each node represents a mesoscopic region of the material of the typical size of one shear transformation~\cite{Puosi_Olivier_Martens_2015_Softmatter11_7639,albaret_2016_PhysRevE93_053002}.
To each site $i$ we associate a local scalar shear stress $\sigma_i$ with a state variable $n_i$, indicating whether the site plastically deforms (${n_i=1}$) or not (${n_i=0}$).
The local stress dynamics is governed by the equation 
\begin{equation}
\partial_t\sigma_i = G_0 \dot{\gamma} + G_0 \sum_{j} \mathcal{G}_{ij} \partial_t \gamma^{\text{pl}}_j,
\label{eq:Eqofmotion} 
\end{equation}
where $G_0$ is the elastic modulus (see illustration in Fig.~\ref{fig:stress-strain}, $\dot{\gamma}$ the externally applied shear rate, ${\partial_t\gamma^{\text{pl}}_j=\frac{n_{j}\sigma_{j}}{G_0\tau_f}}$ the plastic strain rate caused by a rearrangement at site $j$, and ${\tau_f=1}$ a typical stress relaxation time in the fluidized region fixing the time units.
$\mathcal{G}_{ij}$ denotes the Eshelby propagator~\cite{eshelby_1957_ProcRSocLondonA214_376}, that we discretize in Fourier space: ${\hat{\mathcal{G}}(q_i,q_j)=-4 q_i^2 q_j^2/(q_i^2+q_j^2)^2}$.
We apply periodic boundary conditions, since we are interested solely in bulk properties.


A site yields (${n_i=0 \mapsto 1}$) when its stress reaches a local threshold (${\sigma_i \geq \sigma_c^i}$), and recovers its elastic state (${n_i=1 \mapsto 0}$) when a prescribed
local deformation increment is attained after yielding, 
\begin{equation}
\int dt | \partial_t \sigma_i/G_0 + \partial_t\gamma^{\text{pl}}_i| \geq \gamma_c\;.
\label{eq:yieldcriterion} 
\end{equation}
Each time a site yields a new yield stress $\sigma^i_c$ is drawn from a distribution that corresponds to an exponential distribution of energy barriers $E_c=\sigma_c^2/4G_0$ with a lower cut-off in the yield energies.
The parameter choices are exactly the same as in Refs.~\cite{nicolas_martens_barrat_2014_EurPhysLett107_44003, liu_2016_PhysRevLett116_065501}. The only free parameter of the model that we keep is $\gamma_c$, characterizing the relaxation process.
This model has been shown to fit nicely molecular dynamics results on avalanche statistics close to the yielding transition and to produce a flow curve that has a non-trivial exponent of $0.65$ close to yielding transition that crosses over to a region that is well fitted by the $1/2$ HB exponent at intermediate shear rates~\cite{liu_2016_PhysRevLett116_065501}.
Here we restrict ourselves in the study of the model to this second regime.

\begin{center}
\begin{figure}[!t]
\includegraphics[width=1.0\columnwidth, clip]{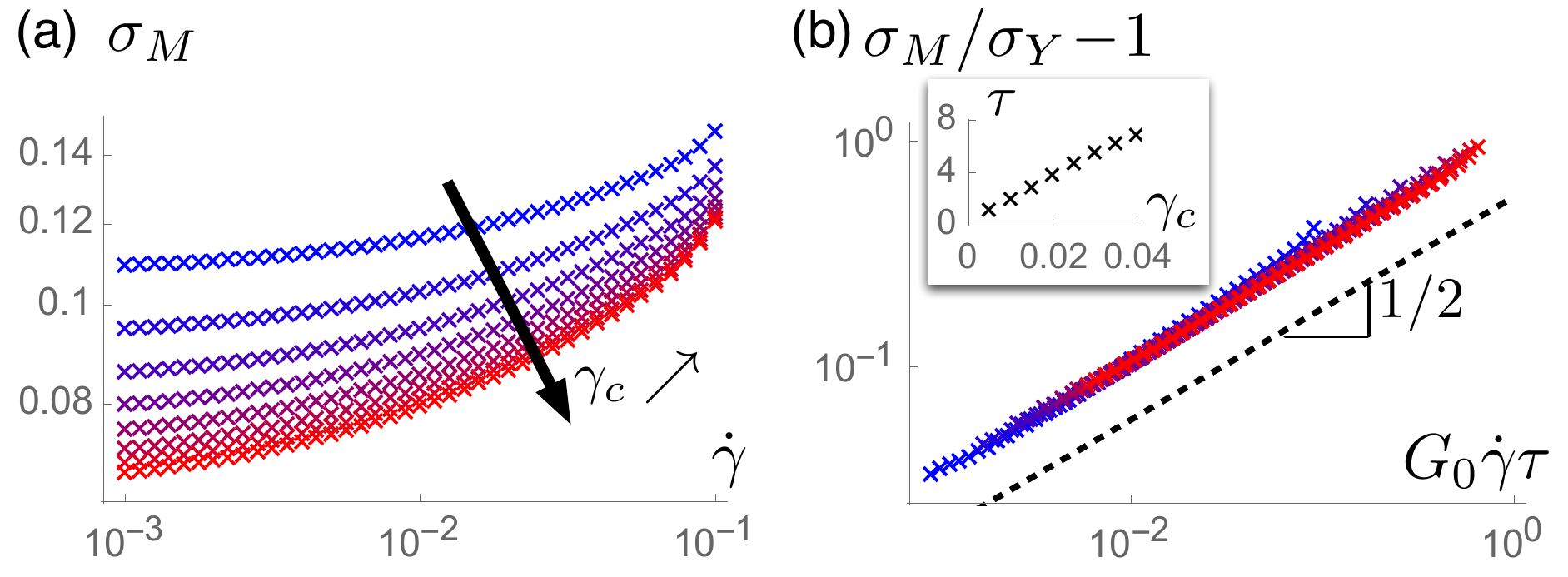}
\caption{{\it Robustness of the 1/2 exponent in the numerical flow-curves -- }
(a)~Steady-state shear stress as a function of the shear rate for different values of the maximal acumulated strain $\gamma_c$ during plastic deformation (see text).
(b)~Shown is $\Delta\sigma/\sigma_y=(\sigma_M-\sigma_y)/\sigma_y$, the rescaled value of the average shear stress subtracting the dynamical yield stress as a function of $G_0\dot{\gamma}\tau$, for different values of $\gamma_c$ in $(0.01 \ldots 0.04)$.
The dashed line of slope $1/2$ is a guide to the eye.
The inset shows the dependence of $\tau$ on $\gamma_c$.}
\label{fig:meso-flowcurves-robustness}
\end{figure}
\end{center}

First we test the robustness of the $1/2$ exponent in this model by analysing the flow curves for several values of $\gamma_c$.
As shown in Fig.~\ref{fig:meso-flowcurves-robustness}(a), the value of the accumulated plastic deformation threshold modifies quantitatively the different features of the flow curve, such as the dynamical yield $\sigma_y(\gamma_c)$. However, we are able to fit all of these curves with an HB law of exponent 1/2 and we are able to extract a time-scale $\tau (\gamma_c)$ from this fit $\sigma_M = \sigma_y(1+(\tau\dot{\gamma})^n)$ as usually done in experiments \cite{cloitre_2003_PhysRevLett90_068303}.
We can thus recover a collapse of these flow curves, by plotting the relative stress value ${(\sigma_M-\sigma_y(\gamma_c))/\sigma_y(\gamma_c))}$ with respect to the shear rate multiplied with $G_0 \tau(\gamma_c)$.
As expected this procedure leads to a collapse of the data onto a curve that is well-fitted by a power law of $1/2$ as shown in Fig.~\ref{fig:meso-flowcurves-robustness}(b).
Moreover, it turns out that the dependence of $\tau$ on $\gamma_c$ is linear in the parameter regime considered, as shown in the inset of Fig.~\ref{fig:meso-flowcurves-robustness}(b).

\begin{center}
\begin{figure}[!t]
\includegraphics[width=\columnwidth, clip]{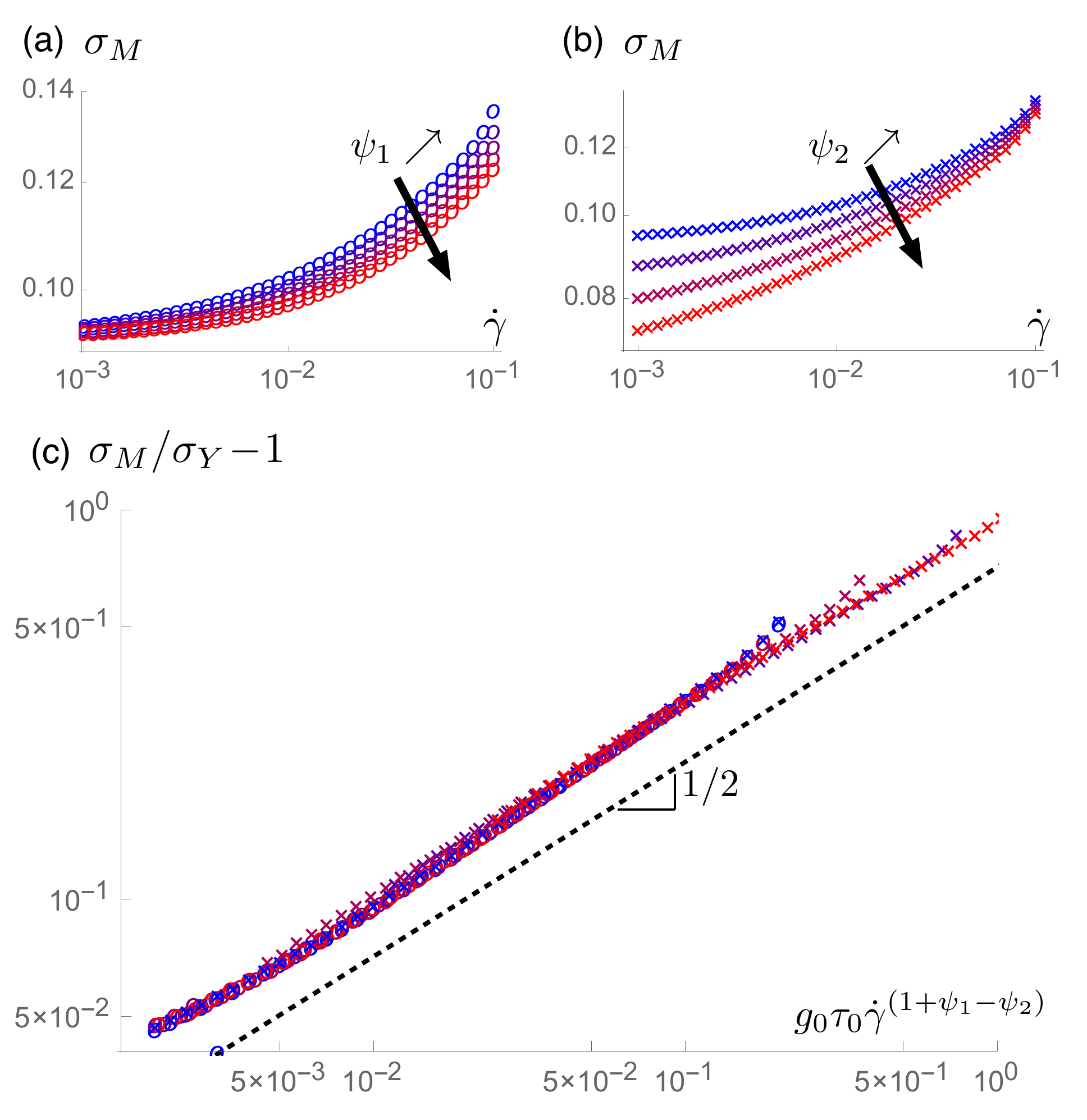}
\caption{{\it Non-trivial exponents in the numerical flow-curves -- }
(a)~Shown is the numerical data of the flow curves obtained from simulations of the spatially resolved mesocopic model with a shear-rate dependent elastic modulus $G_0=\dot{\gamma}^{\psi_1}$ with ${\psi_1 \in \arga{ 0.0,0.05,0.1,0.15,0.2}}$.
(b)~Numerical data for flow curves for simulations with a shear-rate dependent deformation threshold for the local relaxation processes ${\gamma_c=0.01\dot{\gamma}^{-\psi_2}}$ with ${\psi_2 \in \arga{ 0.0,0.05,0.1,0.15}}$).
(c)~Shown is $\Delta\sigma/\sigma_y=(\sigma_M-\sigma_y)/\sigma_y$, the rescaled value of the average shear stress subtracting the dynamical yield stress as a function of the rescaled shear-rate, $g_0\dot{\gamma}\tau_0(\psi_2)(\dot{\gamma})$ of the data presented in (a) and (b).
The fit validates the altered scaling ${\Delta\sigma/\sigma_y\sim\dot{\gamma}^{(1+\psi_1-\psi_2)/2}}$ and the inset shows the flow curve data.}
\label{fig:meso-flowcurves-nontrivial}
\end{figure}
\end{center}

Consequently, regarding the results of the mean-field considerations, we expect that shear-rate dependent elastic properties in the steady state with $G_0=g_0\dot{\gamma}^{\psi_1}$ and shear rate dependent local relaxation processes with $\gamma_c(\dot{\gamma})\sim \dot{\gamma}^{-\psi_2}$ should change the $1/2$ HB exponent to $(1+\psi_1-\psi_2)/2$.
As shown in Fig.~\ref{fig:meso-flowcurves-nontrivial}, this scaling behaviour can be straightforwardly verified in the simulation data of the mesocopic model.


\section{Conclusion}
\label{section-conclusion}

In this study, we propose a coarse-grained elasto-plastic scenario to derive possible physical origins for non-trivial rheological exponents in athermal systems.
In the context of foams and emulsions, earlier studies correlated microscopic properties to the overall flow behaviour, by relating the surface mobility of bubbles or droplets to the shear-rate dependence of the macroscopic viscosity~\cite{denkov_2008_PhysRevLett100_138301}.
However, the full picture clearly needs to take also into account long-range molecular interactions~\cite{cohen-addad_hoehler_2014_CurrentOpinionColloidInterfaceScience19_536}.
Our approach concentrates on this coupling of stress dynamics to elastic response to local rearrangements. It will be interesting to understand relations between the two approaches, as for example the influence of these surface properties on the coarse-grained parameters in the mesoscopic elasto-plastic models.


In this study we discussed how a well-defined HB scaling can be modified by taking into account an effective shear-rate dependence of the elastic modulus, and/or the local relaxation processes,
and we explained why we can observe a HB behaviour with a $1/2$ exponent in many sheared athermal amorphous materials in the framework of ALYS models.
Moreover, we discussed the validity range of this HB behaviour, corresponding to an intermediate shear-rate regime which is bounded
\textit{(i)}~at low shear rates by the development of non-trivial stress fluctuations close to the critical point of zero shear rate~\cite{nicolas_2014_SoftMatter10_4648, lin_wyart_2016_PhysRevX6_011005, liu_2016_PhysRevLett116_065501}, and
\textit{(ii)}~at high shear rates by the crossover towards intermediate scaling regimes depending on the relaxation dynamics and succeded by a completely fluidized, Newtonian regime.
Highlighting furthermore the artefacts that may correspondingly arise when fitting numerical or experimental flow curves, we have thus provided alternative scenarios for measuring non-trivial HB exponents apart from its mean-field predicted value of $1/2$.


We hope to encourage experimental studies and works on particle-based simulations to test our diffusive ALYS scenario, having shown that, in order to assess the existence of a diffusive regime and its shear-rate validity range, it is necessary to combine the characterization of the flow curve, the elastic shear moduli and the typical time scale of the plastic events.
Our analytical predictions are consequently particularly relevant for cases where simultaneous measurements of these quantities are available, e.g.~in foams and emulsions where one has access not only to the frequency dependence of shear and loss moduli~\cite{mason_weitz_1995_PhysRevLett74_1250, liu_1996_PhysRevLett76_3017, dollet_raufaste_2014_CRPhysique15_731,rouyer_2008_EPJE27_309,piau_2007_JNon-NewtonianFluidMech144_1,dinkgreve_2017_RheologicalActa56_189} but also to local observables, like the local stress~\cite{desmond_2015_PhysRevLett115_098302} or the local rheology~\cite{divoux_barentin_manneville_2011_SoftMatter7_9335}.
Superimposing a small amplitude oscillatory motion orthogonal onto steady shear flow should make it possible to directly estimate the shear-rate dependence of the elastic moduli~\cite{jacob_2015_PhysRevLett115_218301} and compare these to the frequency dependence in pure oscillatory experiments.
Another way to estimate this shear-rate dependence would be to measure the linear slopes of the rising parts in the intermittent macroscopic flow response of small sytems~\cite{lauridsen_2002_PhysRevLett89_098303} as a function of the imposed driving rate.
In a recent work on foams~\cite{dollet_bocher_2015_EurPhysJE38_123}, it has been shown that the local elastic deformation is expected to be shear-rate dependent, suggesting that the yielding criteria could also be shear-rate dependent in these materials.


An interesting future aim should also be to understand the effect of the shear-rate dependent parameters on the transient dynamics to make a direct link between our study to the work by Divoux et al.\cite{divoux_barentin_manneville_2011_SoftMatter7_9335} and to further understand the relation between microscopic theories proposed in the literature on foams and the link to effective mesoscopic parameters~\cite{princen_kiss_1989_JColloidInterfaceSci128_176}.

\section{Acknowledgements}

E.~A. acknowledges financial support from the Swiss National Science Foundation under grant No P2GEP2-15586,
from the ERC grant ADG20110209 (`GlassDef'),
in part from the National Science Foundation under grant No.~NSF PHY11-25915,
and from a Simons Foundation grant ($\sharp$454955, Francesco Zamponi).
K.~M. acknowledges financial support of the French Agence Nationale de la Recherche (ANR), under grant ANR-14-CE32-0005 (FAPRES).
We thank Julien Olivier for his invaluable insight on the original HL model,
Eric Bertin for detailed discussions during this work,
and Olivier Dauchot, Benjamin Dollet, Elise Laurenceau, Jean-Louis Barrat, Ezequiel Ferrero and Kamran Karimi for interesting exchanges on this work.




\balance


\bibliographystyle{rsc} 


\providecommand*{\mcitethebibliography}{\thebibliography}
\csname @ifundefined\endcsname{endmcitethebibliography}
{\let\endmcitethebibliography\endthebibliography}{}


\end{document}